\documentclass[11pt,twoside,A4]{article}
\usepackage{times,fancyhdr}
\usepackage[dvips]{graphicx}
\usepackage{latexsym}
\usepackage[affil-it]{authblk}
\usepackage{mathptm}
\usepackage{amsmath}
\usepackage{amssymb}
\usepackage{color}
\usepackage{setspace}
\usepackage{hyperref}
\usepackage{todonotes}
\usepackage[top=4cm, bottom=4cm, left=4cm, right=4cm]{geometry}
\def\beqn{\begin{eqnarray}}
\def\eeqn{\end{eqnarray}}

\begin{document}

\title{Measuring Scientific Broadness}
\author{Tom Price, Sabine Hossenfelder${}^*$}
\affil{\small ${}^*$Frankfurt Institute for Advanced Studies\\
Ruth-Moufang-Str. 1,
D-60438 Frankfurt am Main, Germany
}
\date{}
\maketitle
\begin{abstract}
Who has not read letters of recommendations that comment on a student's
`broadness' and wondered what to make of it? We here propose a way to quantify scientific broadness by a 
semantic analysis of researchers' publications. We apply
our methods to papers on the open-access server arXiv.org
and report our findings. 
\end{abstract}

\section{Introduction}
Most attempts at quantifying scientific output focus on productivity and popularity, measured
by the number of papers, the number of papers in journals with high impact factor,
media mentions, citation counts, or combinations thereof (for a recent review see \cite{Review2015}). This 
focus is problematic not only because it creates perverse incentives \cite{Incentives}, but also because
other criteria fall by the wayside. To identify a suitable candidate for a job opening,
or judge an applicants' qualifications to lead a project to success, other factors
besides productivity and popularity play a role. One of them is the researchers' breadth of knowledge. 
In this present work, we want to propose a simple and efficient way of quantifying
this breadth.

Our aim here is not to argue that any particular level of broadness is good or bad.
Instead, our point of view is that different tasks call for different amounts
of specialization, where here and in the following we will use the word `specialization'
to mean the opposite of `broadness.' We also do not wish to suggest that the
particular measure of broadness which we will propose in the following is the
`right' one. Instead, we merely want to demonstrate that it is a useful 
measure, and one that captures previously unexplored information.

\section{Data}
\label{data}

We based this present analysis on papers from the 
open-access server arXiv, available through the  Open Archives Initiative Protocal for Metadata Harvesting (OAI-PMH) interface \cite{arXivOAI}. The data used for this analysis was downloaded through the interface in February 2018. It contains the metadata of 1,358,923 papers. We use the title, abstract, author's name, date, and arXiv primary category. Before calculating the broadness
values, we also remove all papers with more than 30 authors because we 
expect collaboration papers to be highly specialized by their nature and thus
follow a different distribution. When we analyze the statistical properties of the distribution we further remove all authors with fewer than 20 papers because those
researchers have too few publications to be meaningfully associated with a broadness value. The final sample contains 46,772 authors and 1,350,611 papers.

\section{Analysis}
\label{ana}
 
We analyze the text of the papers in four steps, the details of which will be laid out in the following
subsections. In brief, the procedure works like this:
 
\begin{enumerate}
\item We extract terms from the papers' titles and abstracts. We collect similar terms, such as ``galaxy" and ``galaxies", into clusters which we refer to as ``keywords". We rank each keyword using a combination of how frequently it occurs and the distribution of arXiv primary categories of the papers it appears in. This ranking is based on the assumption that highly generic terms such as ``paper" or ``demonstrate", which make poor keywords, will be more evenly distributed among different arXiv categories. We keep the 40,000 highest-ranking keywords.

\item We create author identifications by matching similar names.

\item We train a statistical model -- latent dirichlet allocation \cite{LDA} -- for the multiset of keywords used by an author. 

\item Once trained, this model allows us to infer a distribution over latent topics for each author. The broadness of an author is then determined as the Shannon entropy of this distribution over topics.

\end{enumerate}

\subsection{Keyword Generation}

We extract the keywords from the titles and the abstracts of papers in our sample. While we could be using pre-existing
classification shemes, such as {\sc MSC} \cite{MSC}, {\sc ACM} \cite{ACM}, or {\sc PACS} \cite{PACS}, this would 
greatly limit the flexibility of our method. The reader be warned that what we refer to as ``keyword'' here is not necessarily a single word, but may be a sequence of words. For example, ``dwarf galaxy'' or ``effective field theory'' would each count as one keyword.

For each paper, we first obtain a sequence of sequences of words by the following steps:
\begin{enumerate}
	\item We concatenate the title and abstract together, with the string ``. " (period and space) in between. We convert the resulting string to lowercase and remove all latex commands.
    \item We obtain a sequence of strings by dividing the string above into contiguous sections. These sections end whenever a period, question mark, open or closed round bracket, open or closed square bracket, semicolon, colon, or comma is encountered.
    \item We break each string in the above sequence of strings into a sequence of words, by dividing it into contiguous sections of characters which contain no whitespace.
\end{enumerate}

We then produce a list which contains all sequences of at most ten words which can be found in the title or abstract of at least 20 papers.
We then remove all entries that begin or end with a stopword, i.e. a word like ``the'' or ``a''.

Next we convert each entry of the list into a reduced form. This we do by removing ``'s'' from the end of every word in a keyword (so ``Einstein field equations" and ``Einstein's field equations' are the same), removing all diacritics, removing all non-alphanumeric characters, and applying the Porter Stemming Algorithm \cite{Porter} to each word that is longer than four characters. Then we join the resulting words of each entry together with no whitespace in between. This means that now, for example, ``noncompact'', ``non-compact',' and ``non compact'' all have reduced form ``noncompact,'' and ``galaxies'' and ``galaxy'' both have the reduced form ``galaxi''. Having done this, we collect sets of terms with the same reduced form. We will henceforth use the term ``keyword" to refer to a set of all terms sharing some common reduced form.  

We now need to identify the keywords that are most relevant.
For this, we define a list $L_p$ for each paper $p$ of keywords that occur in the title or abstract of $p$, and a probability distribution $P(O)$ on keyword occurrences. By a keyword occurence, we mean specifically a triple consisting of an author $a$, a paper $p$ containing that author among its list of coauthors, and an occurrence of a keyword in $p$, or more specifically, an entry of $L_p$. We give the details on the definition of $L_p$ and $P(O)$ in appendix B.

Let $P(C = c)$ be the probability that, in a keyword occurrence randomly selected with probability determined by $P(O)$, the paper's arXiv primary category is $c$. For a keyword $k$, let $P(K = k)$ be the probability that the keyword is $k$, and let $P(C = c \mid K = k)$ be the probability that the category is $c$ given that the keyword is $k$.

We can then define the rank of a keyword $k$ using the Kullback-Liebler divergence $D_{\rm KL}(\mathrm{P}(C \mid K = k) \| \mathrm{P}(C))$ between posterior and prior distributions over arXiv categories, as well as the keyword probability $\mathrm{P}(K = k)$, and a manually chosen constant $r$:

\beqn
\textrm{Rank} = D_{\rm KL}(\mathrm{P}(C \mid K = k) \| \mathrm{P}(C)) \times \left(1 - \exp(-\mathrm{P}(K = k) / r) \right)~,
\eeqn
The effect of $r$ can be roughly summarized as follows: with higher values of $r$, greater precedence is given to commonly used terms. We have found that a value of $r = 4.5 \times 10^{-6}$ (roughly 3 divided by the number of authors in the unfiltered set) gives good results and this value has been used for the following analysis. 
We keep the 40,000 bins that have the highest rank.

Since a keyword then refers to a set of similar terms (such as ``galaxies" and ``galaxy") rather than a single one, we use the keyword's most probable form (as determined by $\mathrm{P}(O)$ mentioned above) as representative.

This completes the generation of the keyword list.

\subsection{Author Identification}

Author names tend to appear in a variety of different forms. For example, a middle name may be included or omitted. A name might be given in full, or only as an initial. There might be an inconsistency in the usage of diacritics. We therefore use the following
procedure to collect names which likely refer to the same person.

Note that, in the following process, each name must consist of at least two words. We ignore all authors whose name, as given by the arXiv data, consists of only one word.

First, we normalize each name by removing all periods and commas, converting it to lower case, and removing all diacritics. Let $N$ denote the set of all the normalizations of the names encountered.

Next, we create a binary relation, $\sim$, that measures the compatibility of two name parts $p_1$ and $p_2$, where $p_{\rm i}$ is either
a first, middle, or last name but not combinations thereof.  
We call two name parts compatible, $p_1 \sim p_2$,  if they are identical or one is just the initial of the other.

Using this, we define another relation, $\approx$, for  two full names $n_1$ and $n_2$ in $N$, composed of name parts. These names are compatible  -- that is, $n_1 \approx n_2$ --  if the last names are identical, the first names are compatible according to $\sim$, and at least one of the following two conditions hold:

\begin{enumerate}
	\item At least one of the two names has no middle names given.
    \item Each name has the same number of middle names given, and each middle name from one is compatible with the corresponding middle name from the other.
\end{enumerate}

The relation $\approx$ is not an equivalence relation: it is reflexive and symmetric, but not transitive. However, we can create an equivalence relation, $\equiv$, from $\approx$. We start with defining $\equiv$ as equal to $\approx$, but whenever we have a failure of transitivity of $\approx$, say $n_1 \approx n_2$ and $n_2 \approx n_3$ but not $n_1 \approx n_3$, we remove $n_2 \equiv x$ and $x \equiv n_2$ for all possible $x$. In other words, every name that is in the middle of some failure of transitivity loses all its neighbors. The resulting relation $\equiv$ then must be an equivalence relation and its equivalence classes are what we will use as author identifiers. 

The author identification leaves us with 664,057 authors. 

\subsection{Creating the LDA model}

Latent Dirichlet Allocation ({\sc LDA}) is a way of generating a probability distribution for a collection of documents. Here, a document is a sequence of words, and a word is an element of a finite set referred to as `the vocabulary'.  {\sc LDA} works by representing the documents as 
mixtures of `latent' topics, and then characterizes these topics by a distribution over
words \cite{LDA}. 

To apply {\sc LDA} to the set of arXiv authors, we take the vocabulary to be our list of 40,000 keywords. Each `document' corresponds to an author, and the sequence of words within each document is the sequence of keywords used in all of the titles and abstracts of that author's papers. To determine the keywords used in each paper and their multiplicities, we use the procedure described in appendix B for creating the keyword lists $L'_p$. These are similar to the keyword lists $L_p$ used in section 3.1, but for the restricted set of the 40,000 highest-ranking keywords.

For training the model, we used the python library Gensim \cite{gensim}. This library uses a training algorithm based on the one described in \cite{onlineLDA}. We used 50 topics, 5 passes, and set the \verb|alpha| and \verb|eta| parameters to \verb|auto|.

Even though there have been many proposed improvements to {\sc LDA} \cite{LDAother4,LDaother3,LDAother2,LDAother1} we decided to use {\sc LDA} because it is widely recognized and there exist implementations in popular open source libraries.

\subsection{Measuring Broadness}

With the trained {\sc LDA} model, we can compute the joint probability density $p(\theta, \mathbf{z}, \mathbf{w})$ of a probability distribution over latent topics $\theta$, a sequence of topics $\mathbf{z}$, and a sequence of keywords $\mathbf{w}$. In principle, given a sequence of keywords $\mathbf{w}_a$ used by an author $a$, we can obtain a single probability distribution over topics for this author by taking the expected value of $\theta$ given $\mathbf{w}_a$. 

Computing this value is intractable in general \cite[Section 5.1]{LDA}. However, it is possible to compute an approximation $q(\theta, \mathbf{z})$ to $p(\theta, \mathbf{z} \mid \mathbf{w})$. More specifically, we can choose $q$ to be the probability distribution which minimizes the Kullback-Liebler divergence $D_{\rm KL}(q(\theta, \mathbf{z}) \| p(\theta, \mathbf{z} \mid \mathbf{w}))$ among all probability distributions in a certain family. The details on the definition of this family of probability distributions, and an iterative algorithm for computing $q$, are given in \cite[Section 5.2]{LDA}.

We therefore modify the above definition of the topic distribution of an author in order to make it computationally tractable: instead of taking the expected value of $\theta$ according to the distribution $p(\theta, \mathbf{z} \mid \mathbf{w})$, we take it according to the distribution $q(\theta, \mathbf{z})$. The marginal distribution $q(\theta)$ is given by a Dirichlet distribution, for which there exists a simple explicit formula for the expected value. This operation of determining $q$ and taking the expected value of $\theta$ is performed by the gensim function \verb|LdaModel.getdocumenttopics|. We set the \verb|minimum_probability| parameter to 0, and all other optional parameters kept their default values.

For assigning a topic distribution to an author, we use only their papers with at most 30 coauthors. This is to avoid measuring an author as extremely specialized because they have many papers with a single highly specialized collaboration. We don't apply this filter at any prior stage of the analysis.

We assume that broader authors will have a less predictable topic distribution. The unpredictability of a distribution can be quantified by the Shannon entropy \cite{Shannon}. We therefore define the broadness of an author to be the Shannon entropy of their topic distribution.

\section{Validity}

In this section, we consider the question of whether latent topic entropy is a valid measurement of scientific broadness. To give an affirmative answer to this, we would need to discuss what is meant specifically by “scientific broadness”, for example, by constructing a nomological network \cite{nomological}. We won’t attempt that in this paper, however, we will take steps in the same direction by showing that latent topic entropy has some properties that we would expect a valid measurement of scientific broadness to have, for most reasonable interpretations of “scientific broadness”.

\subsection{Correlations with other broadness metrics}

One way test whether latent topic entropy qualifies as a measure of scientific broadness is by checking the correlation between latent topic entropy and simpler, more direct measurements. To this end, we have measured the correlation between latent topic entropy and two other metrics based on the arXiv primary categories of an author's papers. (The details on these two metrics are given in Appendix A.) 

The first alternative metric, arxiv category entropy, measures how unpredictable the arxiv categories of an author's papers are. The second, which is a measurement of specialization rather than broadness (that is, it should be lower rather than higher for broader authors), is basically how different the category distribution of an author's papers is from the average category distribution of all authors. The correlations are 0.45 and -0.195 respectively, which are both in the same direction that one should expect from the assumption that these are valid measurements of broadness or specialization.

\subsection{Typical keywords of latent topics}

Intertpreting latent topic entropy as a measure of scientific broadness requires the assumption that the latent topics discovered by {\sc LDA} correspond to distinct scientific topics, instead of being, for example, random distributions of unrelated words. We provide  a list of the 20 most common keywords of each latent topic so that the reader can see that we have reason to think this assumption holds true. \footnote{\url{http://lostinmathbook.com/topic\%20keywords.txt}}

\subsection{Consistency}

Even though we use an author's papers in order to determine their latent topic entropy, our intention is to measure an intrinsic property of the author's research style, not a property of a particular set of papers. Hence, if latent topic entropy is a valid measurement of scientific broadness, different subsets of an author's papers should tend to give similar measurements of latent topic entropy. 

We have tested this hypothesis by measuring the correlation between two different latent topic entropy values for each author with at least 40 papers. The first measurement uses a random half of the author's papers (rounded down), and the second measurement uses the remaining papers. We measured a Pearson's $r$ of 0.94 between the two broadness values, indicating that our broadness metric is not very sensitive to the specific set of papers used to compute broadness, as we would hope.

\section{Results}

In this section, we restrict our attention to authors who have at least 20 papers with no more than 30 coauthors. This is so that we have sufficient data to get a meaningful estimate of their broadness.

\label{results}
 
\subsection{Total Population}

\begin{figure}[ht]
 \centering
 \includegraphics[width=.7\textwidth]{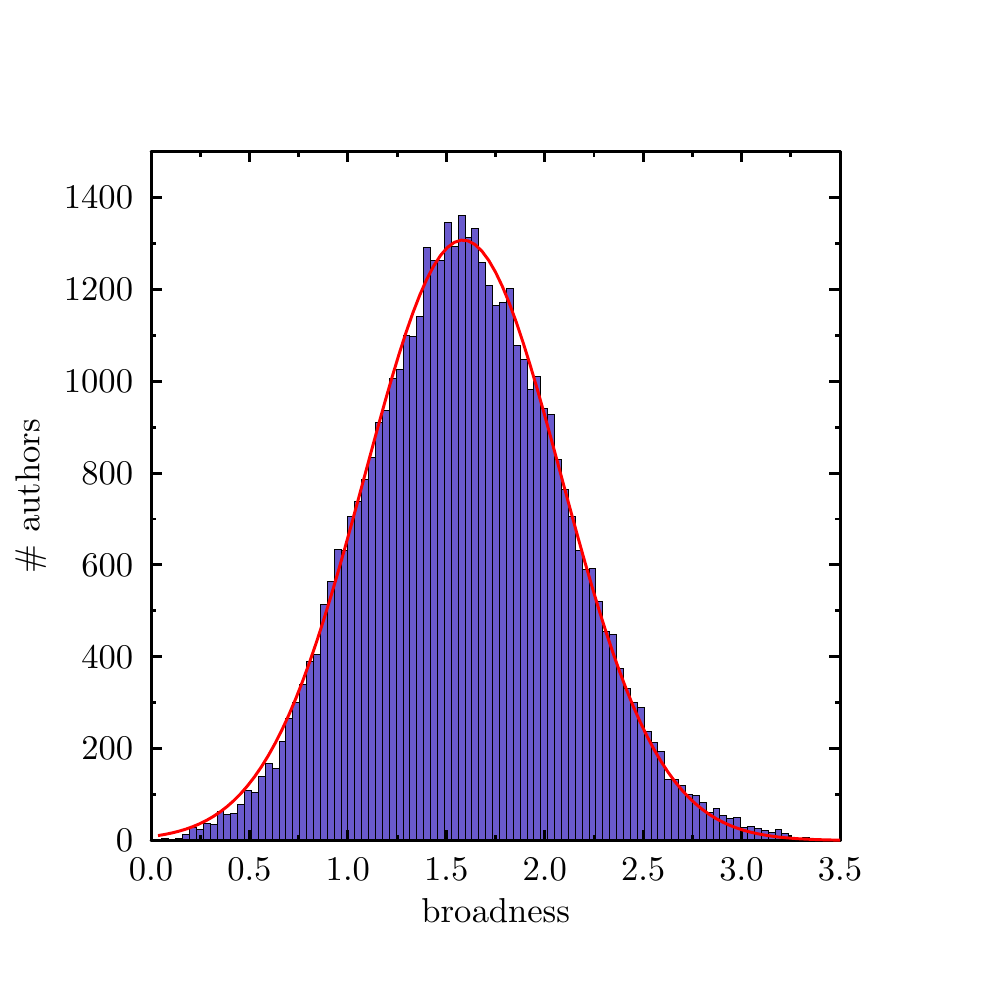}
 \caption{Broadness distribution over authors (blue). Mean value: 1.584, standard deviation: 0.500, total number of authors: 46,772. Normal distribution shown in red.}
 \label{fig:broad}
\end{figure}

In Figure \ref{fig:broad} we depict the distribution of values of
broadness over authors together with a Gaussian fit. The data has a mean value of 1.584
and standard deviation of 0.500. It is close to normal, with a skewness of 0.132 and an excess kurtosis of -0.058.

We note as an aside that if one does not remove papers with more
 than 30 authors (ie keeps papers of large collaborations), the broadness distribution
has a second mode (not shown) which peaks at low broadness. This second mode consists mainly of authors whose papers are mostly with a highly specialized collaboration such as LHC-b or LIGO/VIRGO.

\subsection{ArXiv Categories}

Next we look at authors that are primarily associated with a certain arXiv category, where
we identify an author with a category if it is the primary category of at least 60\% of their papers. Because of the low statistics, we omit categories with fewer than 100
associated authors. In table \ref{tab:cathigh} we list the most broad categories and in table \ref{tab:catlow} we list the least broad categories. The complete list can be downloaded online\footnote{\url{fias.uni-frankfurt.de/~hossi/Physics/author_category.txt}}.

It is instructive to compare these results to the findings of
\cite{Sinatra} which studied (among other things) the frequency by which papers in a sub-field of physics reference the same subfield. In \cite{Sinatra} 
it was found that nuclear physics, astrophysics, the physics of elementary particles and fields, and plasma physics have the highest ratio of self-citations. For the first three of these, there is a tendency for the associated arXiv categories to have low broadness, especially when measured by mean paper broadness. Plasma physics, however, we find to have a high broadness

One possible reason for this discrepancy is that \cite{Sinatra} did not use the arXiv categories, so what they refer to as `plasma physics' is not identical to the category we refer to. Another reason is that broadness just measures a different property to the frequency of self-citations. A category can be broad because its concepts are commonly used also in other categories. This may or may not mean that people who
primarily work in this category commonly refer to papers outside their discipline. 

\begin{table}[ht]
\centering
\begin{tabular}{l||l|l|l}
\textbf{category}  & \textbf{\# authors} & \textbf{mean} & \textbf{standard deviation} \\ \hline \hline
physics.plasm-ph   & 106                 & 1.927         & 0.331             \\ \hline
math.NA            & 113                 & 1.880         & 0.306             \\ \hline
cond-mat.stat-mech & 354                 & 1.870         & 0.332             \\ \hline
math.PR            & 458                 & 1.787         & 0.305             \\ \hline
math-ph            & 181                 & 1.771         & 0.324             \\ \hline
cond-mat.soft      & 281                 & 1.760         & 0.243             \\ \hline
physics.atom-ph    & 164                 & 1.734         & 0.269             \\ \hline
physics.optics     & 231                 & 1.723         & 0.347             \\ \hline
quant-ph           & 1714                & 1.719         & 0.340             \\ \hline
cond-mat.mes-hall  & 1043                & 1.646         & 0.285       
\end{tabular}
\caption{ArXiv categories with the highest mean author broadness. {\label{tab:cathigh}} }
\end{table}

\begin{table}[h]
\centering
\begin{tabular}{l||l|l|l}
\textbf{category} & \textbf{\# authors} & \textbf{mean} & \textbf{standard deviation} \\ \hline \hline
math.GT           & 159                 & 1.293         & 0.276             \\ \hline
cond-mat.str-el   & 912                 & 1.219         & 0.290             \\ \hline
nucl-ex           & 350                 & 1.215         & 0.441             \\ \hline
math.OA           & 109                 & 1.192         & 0.292             \\ \hline
math.GR           & 120                 & 1.180         & 0.296             \\ \hline
astro-ph.CO       & 406                 & 1.179         & 0.378             \\ \hline
hep-th            & 1930                & 1.162         & 0.391             \\ \hline
math.AG           & 407                 & 1.040         & 0.259             \\ \hline
math.RT           & 115                 & 1.008         & 0.327             \\ \hline
astro-ph.GA       & 476                 & 0.920         & 0.409            
\end{tabular}
\caption{{\label{tab:catlow}} ArXiv categories with the lowest mean author broadness.}
\end{table}

Using our trained LDA model, we can also associate a broadness value to a paper in a similar way as for authors. We treat each paper $p$ as an LDA document whose sequence of words is given by the list $L'_p$ (defined precisely in Appendix B) of keywords appearing in the title and abstract. We have calculated the broadness values for arXiv categories as per the average broadness of the papers that have this respective primary category. We omit categories with fewer than 100 associated papers. Since this is a much less restrictive criterion than having at least 100 associated authors, smaller arXiv categories are better represented here. The results are displayed in  tables \ref{tab:cathighpap} and \ref{tab:catlowpap}. The complete list can be downloaded online\footnote{\url{fias.uni-frankfurt.de/~hossi/Physics/paper_category.txt}}.

\begin{table}[ht]
\centering
\begin{tabular}{l||l|l|l}
\textbf{category}  & \textbf{\# papers} & \textbf{mean} & \textbf{standard deviation} \\ \hline \hline
physics.pop-ph    & 781                 & 2.084         & 0.343             \\ \hline
math.HO           & 1568                 & 2.033         & 0.400             \\ \hline
physics.hist-ph   & 1830                 & 2.012         & 0.343             \\ \hline
physics.med-ph     & 1429                & 1.977         & 0.290             \\ \hline
nlin.CG            & 360                 & 1.962         & 0.302             \\ \hline
q-bio.OT           & 406                 & 1.951         & 0.307             \\ \hline
physics.data-an    & 2308                 & 1.944         & 0.353             \\ \hline
physics.class-ph   & 3100                 & 1.939         & 0.352             \\ \hline
patt-sol          & 542                & 1.937         & 0.284             \\ \hline
physics.geo-ph    & 1724                & 1.930         & 0.342            
\end{tabular}
\caption{ArXiv categories with the highest mean paper broadness. \label{tab:cathighpap} }
\end{table}

\begin{table}[]
\centering
\begin{tabular}{l||l|l|l}
\textbf{category} & \textbf{\# papers} & \textbf{mean} & \textbf{standard deviation} \\ \hline \hline
nucl-ex           & 7551                 & 1.414         & 0.438             \\ \hline
math.RT           & 9178                 & 1.406         & 0.419             \\ \hline
astro-ph.EP       & 9758                 & 1.400         & 0.430             \\ \hline
cond-mat.str-el   & 32240                 & 1.373         & 0.390             \\ \hline
astro-ph.HE       & 18998                 & 1.366         & 0.408             \\ \hline
astro-ph.SR       & 25478                 & 1.366         & 0.418             \\ \hline
astro-ph          & 93615                & 1.363         & 0.442             \\ \hline
math.KT           & 1689                 & 1.330         & 0.401             \\ \hline
astro-ph.CO        & 25986               & 1.258         & 0.456             \\ \hline
astro-ph.GA       & 20852                 & 1.160         & 0.447            
\end{tabular}
\caption{ArXiv categories with the lowest mean paper broadness. \label{tab:catlowpap} }
\end{table}

Note that the standard deviations quoted in Tables \ref{tab:cathigh} and \ref{tab:catlow} are for the
distribution in each category. The values do not quantify the deviation of each category's mean value from
that of the entire sample.

For both the mean author broadness and mean paper broadness, applying a one-way ANOVA F-test yields an undetectably small p-value, showing that the differences between categories are exceedingly unlikely to be random fluctuations. 

\subsection{Country broadness}

\begin{figure}[th]
 \centering
 \includegraphics[width=\textwidth]{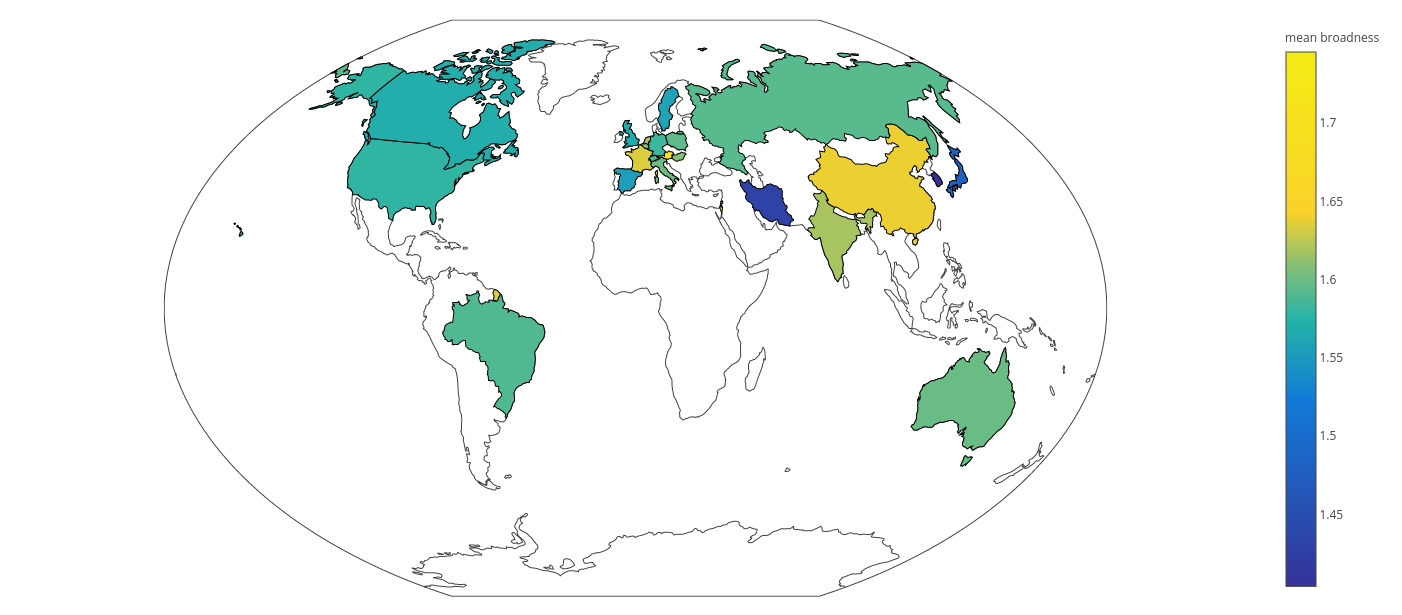}
 \caption{Mean broadness by country}
 \label{fig:country}
\end{figure}

We next quantify the typical broadness per country as the mean broadness of authors in that country. We used the following procedure to associate countries with authors. First, we used arXiv's bulk pdf access \cite{arXivBulk} to download pdf files of arXiv papers up to January 2018. We used Grobid \cite{grobid} to extract the countries of authors from the affiliation data provided in these pdf files. We associated a country with an author if a country was extracted by Grobid for this author in at least one paper, and all countries extracted by Grobid for this author were the same. To get meaningful statistical values, we do not consider
countries which have fewer than 100 associated authors. The results are displayed in Table \ref{tab:country}
and in Figure \ref{fig:country}. The total number of authors here is smaller because
we were not able to link each author to a country, and authors who are linked to countries with fewer than 100 authors in total are not represented.

\begin{table}[ht]
\centering
\begin{tabular}{l||l|l|l}
{\bf Country}                                              & {\bf \# authors} &  {\bf mean} &  {\bf standard deviation} \\ \hline \hline
Israel                                               & 281         & 1.745    & 0.436      \\ \hline
Austria                                              & 127         & 1.705    & 0.398       \\ \hline
China                                                & 998         & 1.639    & 0.496       \\ \hline
France                                               & 1409        & 1.634    & 0.459       \\ \hline
Netherlands                                          & 204         & 1.624    & 0.462       \\ \hline
India                                                & 450         & 1.619    & 0.473      \\ \hline
Belgium                                              & 142         & 1.610    & 0.428       \\ \hline
Hungary                                              & 107         & 1.609    & 0.459       \\ \hline
Italy                                                & 1195        & 1.600    & 0.482       \\ \hline
Australia                                            & 320         & 1.599    & 0.475       \\ \hline
Poland                                               & 301         & 1.595    & 0.428       \\ \hline
Russian Federation                                   & 554         & 1.593    & 0.446      \\ \hline
Brazil                                               & 382         & 1.590    & 0.455      \\ \hline
Switzerland                                          & 197         & 1.590    & 0.425      \\ \hline
Germany                                              & 1397        & 1.583    & 0.451       \\ \hline
United States                                        & 5411        & 1.578   & 0.474      \\ \hline
Canada                                               & 378         & 1.570    & 0.456       \\ \hline
UK and Northern Ireland & 1006        & 1.568    & 0.463        \\ \hline
Sweden                                               & 151         & 1.560    & 0.472        \\ \hline
Spain                                                & 459         & 1.556    & 0.489       \\ \hline
Japan                                                & 1370        & 1.482    & 0.462       \\ \hline
Iran, Islamic Republic of                            & 116         & 1.430    & 0.545       \\ \hline
Korea, Republic of                                   & 181         & 1.404   & 0.438      
\end{tabular}
\caption{\label{tab:country} Mean broadness by country.}
\end{table}

Applying a one-way ANOVA F-test yields a p-value of $3.96 \times 10^{-28}$,
showing that the differences between countries are exceedingly unlikely to be random fluctuations.

We further looked at the correlation between our measure of broadness and
the Nature Index \cite{NatureIndex}. For this we used the weighted fractional count (physical sciences only). The two measures are uncorrelated with a Pearson coefficient of $-0.017$.

\subsection{Gender, career-termination, and $h$-index}

We matched author names with the lists of common female and male names from the 1990 United States Census \cite{names} to identify the gender of an author where possible. This way we were able to identify 6,295 likely male 
and 3,502 likely female authors. (We want to remind the reader that this sample only includes authors with at least 20 papers.) We find small differences
in the mean values and variances of these distributions, but the results are  
not consistent for the four measures of broadness we have tried (see Appendix A). 
We thus conclude that either the gender differences are insignificant or
our present methods do not allow to resolve them.

Next we have analyzed our sample for a correlation between broadness and sudden career terminations. An author is in the terminated-career set if there exists an active period of 10 years in which they have published at least 10 papers, immediately followed by an inactive period, of at least 10 years and extending until the time the data was collected, during which at most 3 papers were published. This is in addition to the usual criterion that they have at least 20 papers with at most 30 coauthors. Our sample contains a total of 1,672 authors with such terminated-careers.

We found that, in the terminated-career set, the mean broadness was 1.483 and the standard deviation was 0.469. We remind the reader that the mean broadness of the whole sample is slightly greater at 1.584, and the standard deviation of the whole sample is 0.5. This is a statistically significant difference in broadness between groups: Welch's t-test gives a p-value of $1.15 \times 10^{-17}$. For all other broadness metrics we investigated (see Appendix A) we also observed that the terminated-career authors were more specialized. The largest p-value obtained was $1.5 \times 10^{-14}$, by the arXiv category Kullback-Liebler divergence metric. 

Therefore, from our analysis, it appears that sudden career terminations are associated with specialized authors. Although the size of the effect on the mean broadness is small, the difference between the means is highly significant. 

We further computed an $h$-index value for each author using the arXiv citation data published by Paperscape \cite{paperscape}. We used the data published in May 2016, which includes citation data up to 2015. Note that the $h$-index value we computed is not necessarily the same as the author's true $h$-index, because the author may not have all their papers on the arXiv. 

We found a Pearson's $r$ value of $-0.183$ between $h$-index and broadness. With all other broadness metrics we tried, we found a slight negative correlation between $h$-index and broadness, except for the arXiv category Kullback-Liebler divergence metric. We suggest a possible explanation for this anomaly in Appendix A.3. Therefore, from this analysis, it appears that there may be a weak positive correlation between specialization and $h$-index, or a weak negative correlation between broadness and $h$-index, respectively.

\subsection{Keyword Broadness}

We can also associate a broadness value to each keyword. For this, we use a probability distribution $P(O')$ on the restricted keyword occurrences $O'$. This is analogous to the distribution $P(O)$, but it uses the restricted set of 40,000 highest-ranking keywords, the restricted set of papers with at most 30 coauthors, and the restricted set of authors with at least 20 papers in the restricted set. The details on this are given in Appendix B.

We can use $P(O')$ to define a broadness value for each keyword: the broadness of $k$ is the expected value of the broadness of the author given that the keyword is $k$.

In table \ref{tab:words}, we list the top ten and bottom ten keywords, subject to the additional restriction that they occur with probability at least $2.1\times 10^{-4}$ (about 10 divided by the size of the restricted set of authors) according to $P(O')$. A complete list can be downloaded online\footnote{\url{fias.uni-frankfurt.de/~hossi/Physics/keywords.txt}}.

We note that the keyword broadness fits
well with the category broadness (Table \ref{tab:cathigh} and \ref{tab:catlow}) in that the most specialized keywords are typical for the astro-ph.X categories and the broadest keywords are typical for many-particle systems found in numerical (math.NA) or probabilistic studies (math.PR) or cond-mat.X applications thereof. 

\begin{table}[ht]
\centering
\begin{tabular}{l||l|l}

                         & \textbf{broadest} & \textbf{most specialized} \\ \hline \hline
\multicolumn{1}{l||}{1}  & agents            & molecular gas                 \\ \hline
\multicolumn{1}{l||}{2}  & chaos             & z = 0       \\ \hline
\multicolumn{1}{l||}{3}  & synchronization   & star-forming                       \\ \hline
\multicolumn{1}{l||}{4}  & chaotic           & star formation rate              \\ \hline
\multicolumn{1}{l||}{5}  & fractal           & early-type galaxies       \\ \hline
\multicolumn{1}{l||}{6}  & sensors           & stellar mass       \\ \hline
\multicolumn{1}{l||}{7}  & network           & z$\sim$2              \\ \hline
\multicolumn{1}{l||}{8}  & memory            & SFR                  \\ \hline
\multicolumn{1}{l||}{9}  & logic             & z$\sim$1                  \\ \hline
\multicolumn{1}{l||}{10} & percolation       & star-forming galaxies     \\ 
\end{tabular}
\caption{Keyword broadness \label{tab:words}}
\end{table}

\section{Conclusion}
\label{conc}
We have proposed and analyzed a new measure to quantify and aggregate research activity
whose purpose is to capture the breadth of a scientist's publications, or their specialization,
respectively. We have found that broadness has little correlation with the $h$-index (of individual authors) or
the Nature Index (of countries), suggesting that it captures
previously unused information. While we do not think that the specific way of measuring
broadness put forward here is the only correct one, we wish to suggest that
broadness is a valuable indicator in particular for nations, institutions, or individuals which strive to improve their interdisciplinary research. 

\section*{Acknowledgements} 

We thank Tobias Mistele for helpful communication. This work was made possible through support by the Foundational Questions Institute (FQXi).

\subsection*{Appendix A: Other Measures}

We tried some other ways to measure broadness before settling on the
the latent topic Shannon entropy used in the main text. For completeness, we here
list other methods that we investigated.

\subsubsection*{1. Kullback-Liebler Divergence}

Instead of measuring the broadness of an author $a$ as the entropy of their topic distribution $T_a$, we could measure it using the Kullback-Liebler divergence $D_{\rm KL}(T_a \| T)$ with the average topic distribution of all authors $T$. We can interpret authors for whom this KL-divergence is low as being broader, and authors for whom it is high as more specialized. The justification for this interpretation is the assumption that a maximally broad author should have a topic distribution equal to $T$, and so the quantity $D_{\rm KL}(T_a \| T)$ measures how different the author's topic distribution is from a maximally broad author.

Note that we can't use $D_{\rm KL}(T \| T_a)$ to define a broadness metric: the Kullback-Liebler divergence is only well-defined if all events that have a probability 0 according to the right distribution also have a probability of 0 according to the left. That is not the case here, since small probabilities in the computed topic distributions often become rounded to 0.

Note that, in general, the entropy of a distribution $P$ is linearly related to the Kullback-Liebler divergence $D_{\rm KL}(P \| U)$ with the uniform distribution $U$ on the same underlying sample space as $P$. From this perspective, we can see that this broadness metric based on Kullback-Liebler divergence is closely related to the main one. The only difference is that the main metric assumes that a perfectly broad author has a uniform topic distribution, while this one assumes that a perfectly broad author has a topic distribution equal to the average topic distribution.

\subsubsection*{2. ArXiv Primary Categories}

Instead of measuring an author’s broadness using on their latent topic distribution, we may use distributions derived from the arXiv primary categories of their papers.

Suppose that the arXiv primary categories of the papers of an author $a$ are sampled from an ideal category distribution $C_a$ for that author, which can be estimated based on the observed categories of this author's papers, but cannot be known. An estimator of the entropy of $C_a$ may be interpreted as a measurement of the author's broadness. Taking the entropy of the maximum-likelihood estimate of $C_a$ (that is, the distribution where the probability of a category is proportional to the number of times it was used in all of the author's papers) is known to be a negatively biased estimator of the true entropy of $C_a$, with the bias becoming less severe as the sample size increases \cite{Basharin}. For example, no matter how broad an author's interests are, if they only have a single paper on arXiv, we will always estimate their category entropy as 0, since every paper of that author is in the same category.

Because of this, we estimated the category entropy of an author by taking a random sample of 20 of their papers without replacement (recall that we restrict our attention to authors with at least 20 papers, so this is always possible), and taking the entropy of the primary category distribution of these 20 papers. This increases the magnitude of the bias of our entropy estimator in most cases, but it becomes more consistent between authors with different numbers of papers, so we avoid systematically measuring a higher broadness value for authors with more papers.

Similarly, we examined another broadness metric obtained by taking the Kullback-Liebler divergence of the category distribution of a 20-paper subset of an author's papers with the average category distribution of all authors. Note that, like the latent topic Kullback-Liebler divergence metric, this is really a measure of specialization since it should decrease for broader authors.

\subsubsection*{3. A comment on $h$-index correlations}
For all the metrics mentioned above, we also measured the correlation with $h$-index like in section 4.4. We found, for all but the arXiv category Kullback-Liebler divergence metric (henceforth refered to as cat-{\sc KLD}), a slight negative correlation between broadness and $h$-index, in agreement with section 4.4. We offer here a possible explanation for why the arXiv category Kullback-Liebler divergence disagreed with the others.

Let $C_a$ be the arXiv category distribution of the 20 randomly-selected papers of some author $a$ used to compute their cat-{\sc KLD}. Let $C$ be the average arXiv category distribution among all authors. The cat-{\sc KLD} metric for the author $a$ is then given by
$$D_{\rm KL}(C_a \| C) = H(C_a, C) - H(C_a)$$

Here, $H(C_a, C)$ is the cross-entropy between $C_a$ and $C$ and $H(C_a)$ is the entropy of $C_a$.

The cross-entropy $H(C_a, C)$ can be interpreted as a measure of how much the author tends to publish in less active arXiv categories. We therefore have that the cat-{\sc KLD} metric will tend to measure authors as more specialized if they publish in less active arXiv categories. This could explain why it correlates negatively with $h$-index (this conflicts with the other metrics, since cat-{\sc KLD} is a measure of specialization and not broadness): the authors with high cat-{\sc KLD} could be receiving fewer citations because they tend to publish in less active categories, where there are fewer authors who might cite their works.

Regarding why the latent topic {\sc KLD} metric doesn't have the opposite correlation with $h$-index for the same reason: while the arXiv categories vary in size by orders of magnitude, the latent topics have relatively consistent average probabilities. Therefore, the cross-entropy term has much less significance in this case.

\subsection*{Appendix B: Details on $L_p$, $P(O)$, $L'_p$, and $P(O')$}

In section 3.1, we describe the rank of a keyword, which quantifies, roughly, a combination of how common the keyword is and how much information it gives about the topic of the paper in question. Our procedure for determining the rank of a keyword depends on the probability distribution $\mathrm{P}(O)$ on keyword occurrences. By a keyword occurrence, we mean specifically a triple consisting of an author $a$, a paper $p$ containing that author among its list of coauthors, and an occurrence of a keyword in $p$, or more specifically, an entry of $L_p$. Here, $L_p$ is a list, possibly with repetition, of the keywords occurring in a paper $p$.

We define the probability $\mathrm{P}(O)$ using the following process:
\begin{enumerate}
	\item Choose an author uniformly at random.
    \item Choose one of this author's papers uniformly at random. Call it $p$.
    \item Choose an entry of $L_p$ uniformly at random.
\end{enumerate}

The probability of a keyword occurrence is then the probability of choosing that author, paper, and entry of $L_p$ in this process.

It remains to give a precise definition of $L_p$ for a paper $p$. For this, we use the following procedure:

\begin{enumerate}
	\item Initialize $S$ as the sequence of sequences of words associated with the paper that is described at the beginning of section 3.1. Initialize $L_p$ as an empty list.
    \item Perform the remaining steps for each nonempty sequence of words in $S$.
    \item If the sequence begins with a keyword, remove the longest possible keyword from the beginning of the sequence (keep in mind that a keyword may contain more than one word, and may contain prefixes that are distinct keywords, such as ``black hole evaporation" and ``black hole"). Add the removed keyword to $L_p$. If the sequence does not begin with a keyword, remove a single word from the beginning.
    \item Repeat the previous step until the sequence is empty.
\end{enumerate}

For each paper, we can also define a restricted list of keywords $L'_p$ (used in section 3.3) in an analogous way, by performing the process above with the restricted set of 40,000 top-ranking keywords instead of the full set. We define the restricted keyword occurrences $O'$ and their distribution $\mathrm{P}(O')$ (used in section 4.5) the same way as $O$ and $\mathrm{P}(O)$, except using $L'_p$ in place of $L_p$, the restricted set of papers with at most 30 coauthors, and the restricted set of authors with at least 20 papers in the restricted set.

\bibliography{broadbib}
\bibliographystyle{unsrt}

\end{document}